\documentclass[preprint,noshowpacs,showkeys,preprintnumbers,amsmath,amssymb,superscriptaddress]{revtex4}

\usepackage{graphicx}
\usepackage{amsmath}
\usepackage{amsbsy}
\usepackage{dcolumn}
\usepackage{bm}
\usepackage{braket}

\begin{document}

\title{
Temperature-dependence of anomalous Hall conductivity in Rashba-type ferromagnets
}

\author{Akimasa Sakuma}
\affiliation{Department of Applied Physics, Tohoku University,
Aoba 6-6-05, Aoba-ku, Sendai 980-8579, Japan}

\date{\today}

\begin{abstract}
We theoretically investigated the anomalous Hall conductivity (AHC) of Rashba-type ferromagnets at a finite temperature, taking into account spin fluctuation. We observed that the intrinsic AHC increases with increasing temperature. This can be understood from the characteristic nature of the spin chirality in the k-space which increases with decreasing exchange splitting (EXS) when the spin-orbit interaction is much smaller than the EXS. The extrinsic part also increases with temperature owing to the enhancement of the scattering strength of electrons due to the thermal fluctuation of the exchange field.
\end{abstract}


\maketitle


\section{Introduction}
Rashba model was originally introduced to express the spin-orbit interactions (SOIs) that occur at the interfaces in asymmetric heterostructures with semiconductors \cite{Rashba}. In recent studies in the field of spintronics, the applicability and usefulness of the Rashba model has been extended by observations, such as spin-orbit torque at the junction interfaces between ferromagnetic metals (FM) and non-magnetic metals (NM) \cite{Manchon,Pesin,Wang,Bijl,Khvalkovskiy,Ueda} and the perpendicular anisotropic magnetoresistance (AMR) effect in heterostructures such as FI/NM \cite{Chen,Lu,Hakayama,Grigoryan,Lin,Zhang2014} or FM/NI (I denotes an insulator) \cite{Zhang2015,Zou}.  In particular, the observations of the perpendicular AMR effect stimulate further interest in the Rashba-type SOI at interfaces, as discussed by recent theoretical studies \cite{Grigoryan,Zhang2015}.  

Also, in the extensive studies on the anomalous Hall effect (AHE) carried out in the last decade \cite{Nagaosa}, the Rashba model including exchange splitting (EXS) has been actively used and have played an important role in clarifying the intrinsic and extrinsic contributions in the anomalous Hall conductivity (AHC) \cite{Inoue,Onoda,Nunner,Ren,Itoh,Sinitsyn2008,Kovalev}. 
Furthermore, the Rashba model with EXS and two-dimensional massive Dirac model \cite{Crepieux,Sinitsyn2007} help us to re-discuss the intrinsic AHE in the context of Berry curvature in k-space \cite{Nagaosa} .
Thus, the Rashba model with EXS is considered not only to play as an effective model for the physical understanding but also to reflect actual bi-layer systems in current spintronics devices. 

We stress here that in AHC a further area of interest is expected to lie in the effects of spin fluctuation at finite temperature when the Rashba SOI dominates the AHE.  It has been widely accepted that, in bulk ferromagnetic systems, AHC generally decreases with increasing temperature as shown experimentally \cite{Zeng,Ye2012}.  In bulk ferromagnetic systems where the EXS ($z$-direction) is much larger than the spin-orbit splitting, AHE is mainly governed by the $l_zs_z$ part in intra-atomic SOI and then the existences not only of $\langle s_z\rangle$ but also of $\langle l_z\rangle$ substantiate the AHE. This would result in decreasing behaviour with increasing temperature. In Rashba-type ferromagnets, on the other hand, only the spin flip terms ($\sigma_x$, $\sigma_y$) exist in the SOI and then the dependence of Berry cuvature (spin chirality in k-space) on the magnitude of EXS toward $z$-direction is not so simple. Therefore,the effects of spin fluctuation on the AHC at finite temperatures are expected to be different from those in the usual transition metals.

Motivated by this peculiar situation of the Rashba model, in this work, we investigated the AHC of Rashba-type ferromagnets at finite temperatures using the tight-binding lattice model, considering spin fluctuations in the disordered local moment (DLM) scheme. The lattice model is not only realistic but also enables us to study finite-temperature magnetism because the theoretical realization of the magnetic phase transition requires a finite-band-width model. 
The most distinctive feature of the AHC that we observed using the Rashba model is the increase of the intrinsic AHC with increasing temperature. This can be understood in terms of the spin chirality in k-space which increases with decreasing EXS when the Rashba SOI is much smaller than the EXS. Although, such a behaviour has not yet been observed experimentally, we suggest that the physical picture found in this work might lurk in an AHE in Rashbe-type ferromagnets.

The structure of this paper is as follows. In Section 2, we introduce a one-electron Hamiltonian to describe  Rashba-type ferromagnets in the tight-binding lattice model and express the AHC by using Kubo-Streda formula \cite{Kubo} within the framework of the coherent potential approximation for the spin configuration. In Section 3, we present numerical calculation results for the temperature dependence of the magnetization and the AHC and provide some relevant discussion. Finally, we summarize our findings in Section 4.

\section{Model and calculation method}
Focusing on an interface of a bi-layer system where Rashba SOI appears, we consider, for simplicity, a two-dimensional square lattice with a lattice constant $a$ in order to calculate the AHC in the interfacial Rashba layer.  
The Rashba Hamiltonian in the tight-binding lattice can be described by~\cite{Ando}
\begin{align}
  H_{\rm Rashba} =& -2t\sum_{\bm k,\sigma}(\cos{ak_x}+\cos{ak_y})
  n_{\bm k,\sigma}\notag  \\
 & + \lambda\sum_{\bm k,\sigma,\sigma'}
  [(\sigma_x)_{\sigma,\sigma'}  \sin{ak_y}-(\sigma_y)_{\sigma,\sigma'}\sin{ak_x}]
   c_{\bm k,\sigma}^{\dagger}c_{\bm k,\sigma'} ,\label{eq.1}
\end{align}
where $n_{\bm k,\sigma}=c_{\bm k,\sigma}^{\dagger}c_{\bm k,\sigma}$ and $c_{\bm k,\sigma}^{\dagger}(c_{\bm k,\sigma})$ denotes the annihilation (creation) operator of electrons with a momentum $\bm k$ and spin $\sigma$ . The first term represents the two-dimensional hopping term with strength $t$ and the second term represents the Rashba-type SOI in the tight-binding scheme with coupling constant $\lambda$. In principle, to study the finite-temperature magnetism of an itinerant electron system, it is necessary to consider the Coulomb interaction between electrons. The typical approach when examining such a system is the functional integral method to perform the Hubbard-Stratonovich transformation~\cite{Hubbard,Stratonovich}. Accordingly, we can address a single-particle system in the auxiliary fields $\bm\Delta_i(\tau)$ (magnetic texture) that fluctuate in time and space $i$ (lattice site). If we use the saddle-point approximation in terms of the magnitude of $\bm\Delta_i(\tau)$, the remaining degree of freedom is the direction $\bm e_i(\tau)$ of the fields. Here, we can regard the auxiliary field as an exchange field defined by $\Delta_{\rm ex}\bm e_i(\tau)$. Furthermore, the adiabatic approximation can be naturally introduced in the thermally fluctuating field, which leads to $\bm e_i(\tau) \Rightarrow\bm e_i $. This treatment is the so-called DLM scheme. Using these approximations, the effective Hamiltonian under a certain configuration of exchange field can be written as
\begin{align}
  H\{\bm e\} = H_{\rm Rashba} - \Delta_{\rm ex}\sum_i\bm e_i\cdot\bm {\sigma}_i,\label{eq.2}
\end{align}
where $\bm {\sigma}_i = \sum_{\sigma,\sigma'}(\bm {\sigma})_{\sigma,\sigma'}
c_{i,\sigma}^{\dagger}c_{i,\sigma'} $  with $\bm {\sigma}$  being the Pauli matrix. Here $\{\bm e\}$ implies the spatial configuration of exchange field directions whose degree of randomness is determined depending on temperature using the functional integral method.

The Hamiltonian (eq.(\ref{eq.2})) describes the two-dimensional system where the ferromagnetism is unstable if it could stand alone.  However, we consider here an interfacial layer of a bi-layer system where the ferromagnetic state is stably sustained with a finite thickness.  Moreover, the presence of SOI in this layer gives rise to magnetic anisotropy energy, which would further stabilize the ferromagnetism at finite temperature. These situations may permit us to adopt the single-site approximation to express the thermally fluctuating spins as scattering centres for electrons,  which can usually be dealt with using the coherent potential approximation (CPA) \cite{Cyrot}. If time dependency is included, this becomes the dynamical mean field theory.  The CPA condition within the functional integral method is given by 
\begin{align}
  &\Bar{G}(\epsilon_+) =\frac{1}{N}\sum_{\bm k}\Bar{G}_{\bm k}(\epsilon_+)
    =\frac{1}{N}\sum_{\bm k}[\epsilon_+ - (H_{\rm Rashba})_{\bm k}-\Sigma({\epsilon_+})]^{-1}, \tag{3a}\label{eq.3a}\\
  &\braket{t(\bm e)}_{\bm e}
    =\langle (-\Delta_{\rm ex} \bm e\cdot\bm {\sigma}-\Sigma(\epsilon_+))
   [1-\Bar{G}(\epsilon_+)(-\Delta_{\rm ex} \bm e\cdot\bm{\sigma}-\Sigma(\epsilon_+))]^{-1}\rangle_{\bm e}=0, \tag{3b}\label{eq.3b}\\
  & \braket{\cdots}_{\bm e}\equiv \int w(\bm e)\cdots d\bm e, \tag{3c}\label{eq.3c}\\
  & w(\bm e)=e^{-\Omega(\bm e)/k_{\rm B}T}/\int e^{-\Omega(\bm e)/k_{\rm B}T}d\bm e, \tag{3d}\label{eq.3d}\\
  & \Omega(\bm e)=-(1/\pi){\rm Im}\int_{-\infty}^{\infty}d\epsilon f(\epsilon)
      {\rm Tr}_\sigma\ln [1-(-\Delta_{\rm ex}\bm e\cdot\bm{\sigma}-\Sigma(\epsilon_{+}))\Bar{G}(\epsilon_+)], \tag{3e}\label{eq.3e}
\end{align}
where $\Bar{G}(\epsilon_+) $ and $\Sigma(\epsilon_+)$ are the coherent Greens function and the coherent potential, respectively, with $\epsilon_+\equiv \epsilon+i\delta$ where $\delta$ is the infinitesimal positive value. $N$ and $f(\epsilon)$ are the number of unit cells and the Fermi distribution function, respectively. The quantity denoted by $t(\bm e)$ indicates a $T$-matrix with the scattering potential 
$-\Delta_{\rm ex}\bm e\cdot\bm{\sigma}-\Sigma(\epsilon_{+})$ at a certain site and its configurational average 
$\braket{t(\bm e)}_{\bm e}$ must vanish to satisfy the CPA condition. The configurational average is calculated with the possibility weight $w(\bm e)$ of the exchange field having direction $\bm e$ at a certain site.  

The AHC is given by the so-called Kubo-Streda formula \cite{Kubo}
\begin{align}
 & \sigma_{xy}=\sigma_{xy}^{\rm I}+\sigma_{xy}^{\amalg}, \tag{4a}\label{eq.4a} \\
 & \sigma_{xy}^{\rm I}=\frac {\hbar}{4\pi\Omega}\int _{-\infty}^{\infty}d\epsilon (-\frac{\partial f(\epsilon)}{\partial \epsilon})
    {\rm Tr}\braket{J_x(G^+-G^-)J_yG^- - J_xG^+J_y(G^+-G^-)}_{\{\bm e\}},\tag{4b}\label{eq.4b}\\
 & \sigma_{xy}^{\amalg}=\frac {\hbar}{4\pi\Omega}\int _{-\infty}^{\infty}d\epsilon f(\epsilon)
    {\rm Tr}\langle J_x\frac{dG^-}{d\epsilon}J_yG^- -J_xG^- J_y\frac{dG^-}{d\epsilon}
    -J_x\frac{dG^+}{d\epsilon}J_yG^+ + J_xG^+ J_y\frac{dG^+}{d\epsilon}\rangle_{\{\bm e\}},\tag{4c}\label{eq.4c}
\end{align}
where $\Omega$ denotes the system area given by $\Omega=Na^2$ , and $G^\pm\equiv(\epsilon_\pm-H{\{\bm e\}})^{-1}$. The electric current operators are expressed by
\begin{align}
 J_{x(y)}&=\frac{e}{\hbar}\frac{\partial H\{\bm e\}}{\partial k_{x(y)}} \notag \\
              &=\frac{2tae}{\hbar}\sum_\sigma\sin{ak_{x(y)}}c_{\bm k,\sigma}^{\dagger}c_{\bm k,\sigma}
              -(+)\frac{\lambda ae}{\hbar}\sum_{\sigma ,\sigma'}(\sigma_{y(x)})_{\sigma,\sigma'}\cos{ak_{x(y)}}
              c_{\bm k,\sigma}^{\dagger}c_{\bm k,\sigma'}. \tag{5}\label{eq.5}
\end{align}
The average $\braket{\cdots}_{\{\bm e\}}$ in eqs.(4) represents the configurational average in the exchange field direction, which, at this stage, is different from the single-site scheme $\braket{\cdots}_{\bm e}$ in eq.(\ref{eq.3c}). The first and the second terms in eq.(\ref{eq.4a}) are the so-called Fermi surface and Fermi sea terms, respectively. The Fermi surface term consists of a type of ${\rm Tr}\braket{J_xG^aJ_yG^b}_{\{\bm e\}}$, where $a, b$  denote either $+$  or $-$. This can be expanded in the single-site scheme using the coherent Green functions as~\cite{Turek} 
\begin{align}
 {\rm Tr}\braket{J_x G^a J_y G^b}_{\{\bm e\}}
={\rm Tr}_\sigma \sum_{\bm k} J_x \Bar{G}_{\bm k}^a J_y \Bar G_{\bm k}^b
+(1/N)\sum_{L,L'}[A_x^{b,a}]_L[\Gamma^{a,b}]_{L,L'}[A_y^{a,b}]_{L'}, \tag{6}\label{eq.6}
\end{align}
where $L\equiv(\sigma_1, \sigma_2)$  and
\begin{align}
& [A_{\alpha}^{a,b}]_L=\sum_{\bm k} [\Bar G_{\bm k}^a J_\alpha \Bar G_{\bm k}^b]_{\sigma_1,\sigma_2}, \tag{7a}\label{eq.7a}\\
& [\Gamma^{a,b}]_{L,L'}=[\gamma^{a,b}]_{L,L'}+\sum_{L"}[\gamma^{a,b}\chi^{a,b}]_{L,L"} [\Gamma^{a,b}]_{L",L'} 
=\sum_{L"}[(1-\gamma^{a,b}\chi^{a,b})^{-1}]_{L,L"} [\gamma^{a,b}]_{L",L'},\tag{7b}\\
& [\chi^{a,b}]_{L,L'}=\frac{1}{N}\sum_{\bm k}[\Bar G_{\bm k}^a]_{\sigma_1,\sigma_1'} [\Bar G_{\bm k}^b]_{\sigma_2',\sigma_2}
    -[\frac{1}{N}\sum_{\bm k}\Bar G_{\bm k}^a]_{\sigma_1,\sigma_1'}[\frac{1}{N}\sum_{\bm k}\Bar G_{\bm k}^b]_{\sigma_2',\sigma_2},\tag{7c}\\
& [\gamma^{a,b}]_{L,L'}=\braket{[t^a(\bm e)]_{\sigma_1,\sigma_1'}[t^b(\bm e)]_{\sigma_2',\sigma_2}}_{\bm e},\tag{7d}\\
& [\gamma^{a,b}\chi^{a,b}]_{L,L'}=\sum_{L"}[\gamma^{a,b}]_{L,L"} [\chi^{a,b}]_{L",L'}\notag \\
&=\langle\frac{1}{N}\sum_{\bm k}[t^a(\bm e)\Bar G_{\bm k}^a]_{\sigma_1,\sigma_1'} [\Bar G_{\bm k}^bt^b(\bm e)]_{\sigma_2',\sigma_2}\rangle_{\bm e} 
-\langle[t^a(\bm e)\frac{1}{N}\sum_{\bm k}\Bar G_{\bm k}^a]_{\sigma_1,\sigma_1'}
               [\frac{1}{N}\sum_{\bm k}\Bar G_{\bm k}^bt^b(\bm e)]_{\sigma_2',\sigma_2}\rangle_{\bm e}.\tag{7e}
\end{align}
Here, we redefine the coherent Green function $\Bar G_{\bm k}^\pm(\epsilon)\equiv\Bar G_{\bm k}(\epsilon_\pm)$  in eq.(\ref{eq.3a}) and $t^\pm(\bm e)$ by using $\Bar G_{\bm k}^\pm(\epsilon)$ and $\Sigma^\pm(\epsilon)\equiv\Sigma(\epsilon_\pm)$ in eq.(\ref{eq.3b}). The first term of eq.(\ref{eq.6}) is the coherent term $\sigma_{xy}^{\rm {I\,int}}$ that describes the intrinsic contribution and the second term is the vertex correction that represents the extrinsic part $\sigma_{xy}^{\rm {I\,ext}}$, including the skew scattering and 
side-jump terms \cite{Sinitsyn2007,Onoda,Nunner}. Note that the vertex terms are constructed of $T$-matrices and composed of scattering centres of both non-Gaussian and Gaussian distributions that correspond to the skew and intrinsic skew scattering contributions, respectively \cite{Nunner,Czaja}. The Fermi sea term $\sigma_{xy}^{\amalg}$ constitutes of terms like ${\rm Tr}\braket{J_x (\partial G^a/\partial \epsilon) J_y G^a}_{\{\bm e\}}$, which can be expanded as
 \begin{align}
{\rm Tr}\langle J_x \frac{\partial G^a}{\partial \epsilon} J_y G^a\rangle_{\{\bm e\}}
=-{\rm Tr}_\sigma \sum_{\bm k} J_x \Bar G_{\bm k}^a\Bar G_{\bm k}^a J_y \Bar G_{\bm k}^a
+{\rm Tr}_\sigma \sum_{\bm k}J_x \Bar G_{\bm k}^a\frac{\partial\Sigma^a}{\partial\epsilon}
\Bar G_{\bm k}^a J_y \Bar G_{\bm k}^a.\tag{8}
\end{align}
By adopting the theoretical work for the AHC under the CPA \cite{Turek}, the matrix $\partial\Sigma^a/\partial\epsilon$ can be expressed as
\begin{align}
&\left[\frac{\partial\Sigma^a}{\partial\epsilon}\right]_L=[\kappa]_L+\sum_{L'}[\gamma^{a,a}\chi^{a,a}]_{L,L'}\left[\frac{\partial\Sigma^a}{\partial\epsilon}\right]_{L'}=\sum_{L'}[(1-\gamma^{a,a}\chi^{a,a})^{-1}]_{L,L'} [\kappa^{a,a}]_{L'},\tag{9}\\
&[\kappa^{a,a}]_{L}=-\sum_{L'}\langle\frac{1}{N}\sum_{\bm k}[t^a(\bm e)\Bar G_{\bm k}^a]_{\sigma_1,\sigma_1'}
[\Bar G_{\bm k}^at^a(\bm e)]_{\sigma_2',\sigma_2}\rangle_{\bm e}\delta_{\sigma_1',\sigma_2'}.\tag{10}
\end{align}
Note that in $\sigma_{xy}^{\amalg}$, the current vertex correction term vanishes when the current operator is defined with an inter-site hopping, as in eq.(\ref{eq.5}), and $a=b$  in eq.(\ref{eq.7a}) \cite{Turek}. The diagrammatic representations of $\Gamma^{a,b}$ and $\partial\Sigma^a/\partial\epsilon$ are shown in Fig. 1. 

\begin{figure}
\centering
\includegraphics[width=0.6\textwidth]{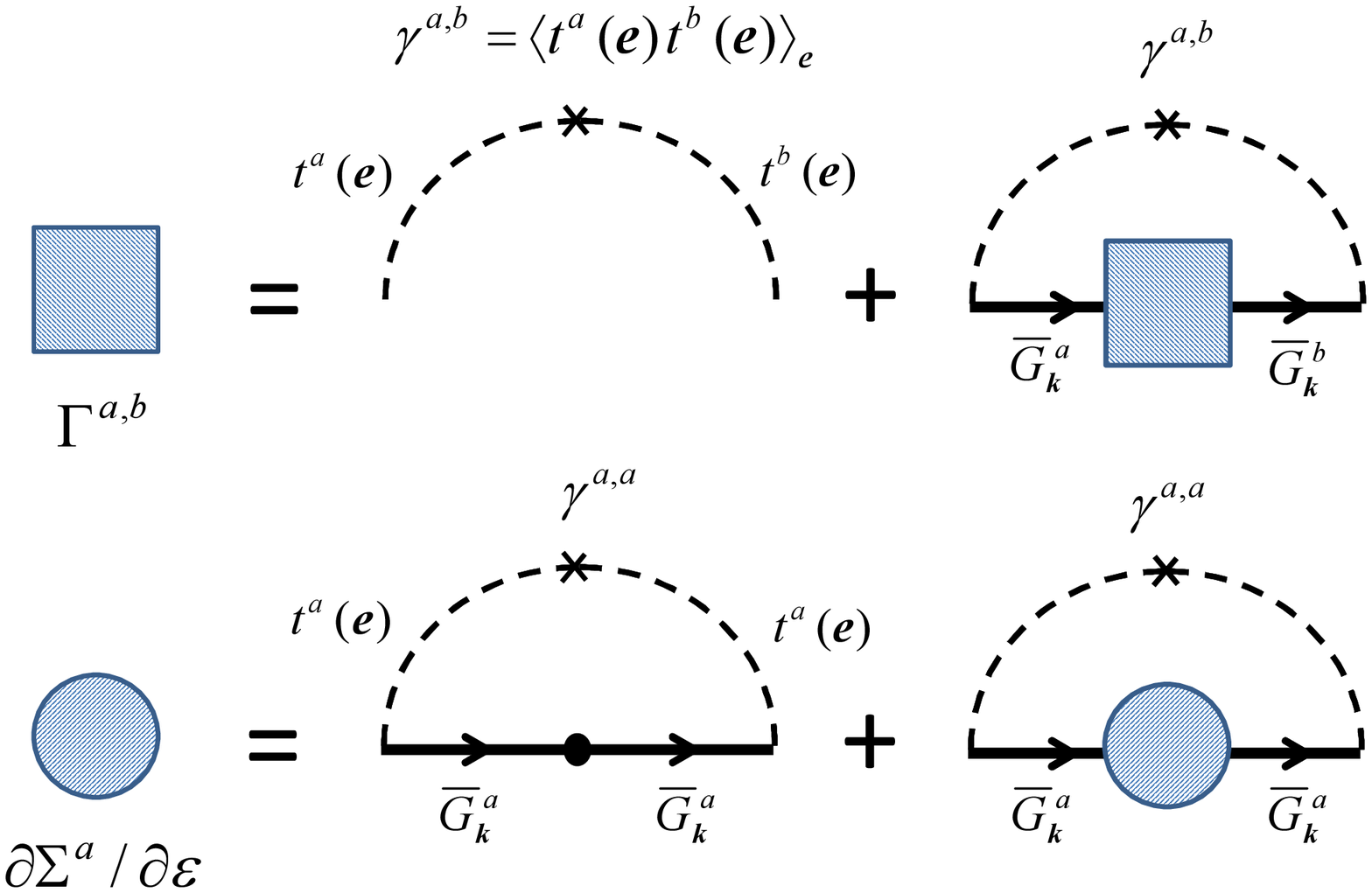}
\caption{Diagrammatic representations of (a) $\Gamma$ and (b) $\partial\Sigma^a/\partial\epsilon$.\\ 
The solid lines represent coherent Green functions and the dashed lines the $T$-matrices due to spin fluctuation denoted by crosses.}
\end{figure}

\section{Results and discussion}
For the numerical calculations, we first set the parameters $\Delta_{\rm ex}=0.5$ and $\lambda=0.01$ in a 
unit of $W=2t$ which we consider to be reasonable for ferromagnetic metals. Under these parameters, we must carefully chose the electron concentration $n$, because magnetism is mainly governed by $n$ or the Fermi level position as shown in Ref.\cite{SakumaJPSJ}. To determine $n$, it is convenient to evaluate the effective exchange constant $J_0$ as a function of Fermi energy $\epsilon_{\rm F}$ at $T=0$. This is defined by \cite{Liechtenstein}
 \begin{align}
J_0(\epsilon_{\rm F})=\sum_{i(\neq0)}J_{0,i}=-\frac{1}{4\pi}{\rm Im}\int_{-\infty}^{\epsilon_{\rm F}}d\epsilon
[2\Delta_{\rm ex}(G^\uparrow(\epsilon_+)-G^\downarrow(\epsilon_+))+(2\Delta_{\rm ex})^2 
G^\uparrow(\epsilon_+)G^\downarrow(\epsilon_+)],\tag{11}\label{eq.11}
\end{align}
where $G^\sigma(\epsilon_+)(\sigma=\uparrow,\downarrow)$  is the local Green function (at 0-th site) of a provisional ferromagnetic state, 
\begin{align}
G^\sigma(\epsilon_+)=\frac{1}{N}\sum_{\bm k}[\epsilon-(H_{\rm Rashba})_{\bm k} +\Delta_{\rm ex}\sigma_z+i\delta]_{\sigma,\sigma}^{-1}.\tag{12}\label{eq.12}
\end{align}
Throughout this study, the infinitesimal value was set at $\delta=0.001W$. Here, we choose $z$-direction as the exchange splitting direction $\Delta_{\rm ex}\sigma_z$, for instance, in order to realize AHE in the x-y plane. Note here that $J_0$ is different from $\Delta_{\rm ex}$. The $\Delta_{\rm ex}$ reflects intra-atomic exchange splitting (Hund coupling) while $J_0$ denotes the total exchange fields ($\sum_{i(\neq0)}J_{0,i}$) acting on a certain site (0-th site) from surrounding sites; a positive $J_0$ indicates that the magnetic moment at the 0-th site is forced to align parallel (ferromagnetic) to that of surrounding sites at the ground state, while a negative value suggests that the ferromagnetic configuration is unstable. In Fig. 2, we show $J_0(\epsilon)$ together with the density of states (DOS) projected into each spin state. $J_0(\epsilon)$ implies a variation of $J_0$ when one moves the Fermi level position with the variable of $\epsilon$. One can see that $J_0(\epsilon)$ exhibits negative value around the region of half-filling ($\epsilon\sim0$), which is a natural feature of $J_0(\epsilon)$ as shown in Ref.\cite{SakumaIEEE}.  Based on this behaviour, we choose $n=0.25$ to realize the ferromagnetic state ($J_0>0$) at the ground state, where both the upper and lower branches have a finite DOS, as indicated by the arrow in Fig. 2.  Furthermore, we found by evaluating the magnetic anisotropy constant (not shown here) that the easy direction of magnetization is $z$-direction when $n=0.25$. This situation is consistent with the initial setting ($\Delta_{\rm ex}\sigma_z$ in eq.(\ref{eq.12})).  The details of the theoretical analysis on the magnetic anisotropy in the Rashba-type ferromagnets will be presented elsewhere. Turning to the DOS, the shape is almost the same as that of the usual two-dimensional square lattice model, since the exchange splitting $\Delta_{\rm ex}$ is much larger than $\lambda$. As a reference for comparison, we calculate the case for $\Delta_{\rm ex}=0.1$ and $\lambda=0.5$, which may not be a realistic ferromagnetic system. In such case with $\lambda>2\Delta_{\rm ex}$, the lower branch of the energy dispersions exhibits a double minimum and a sharp peak at the band edge in the DOS, as shown in Fig. 3.
 
\begin{figure}
\centering
\includegraphics[width=0.7\textwidth]{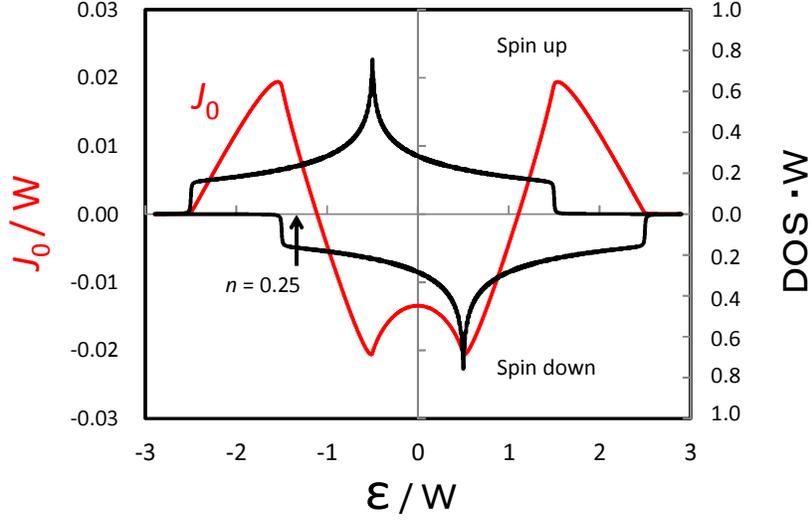}
\caption{Density of states (DOS) at $T=0$ projected into up and down spin states and the effective exchange constants $J_0$ given by eq. (\ref{eq.11}), with the parameters $\Delta_{\rm ex}=0.5$ and $\lambda=0.01$ in unit of 
$W=2t$. The arrow indicates the Fermi level settled in this work, corresponding to the electron concentration of $n = 0.25$.}
\end{figure}

\begin{figure}
\centering
\includegraphics[width=0.7\textwidth]{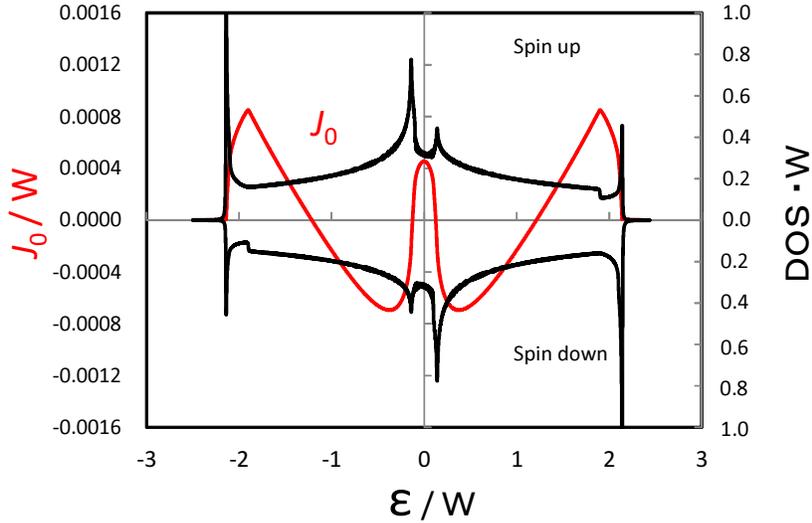}
\caption{Density of states (DOS) at $T=0$ projected into up and down spin states and the effective exchange constants $J_0$ given by eq. (\ref{eq.11}), with the parameters $\Delta_{\rm ex}=0.1$ and $\lambda=0.5$ in unit of 
$W=2t$.}
\end{figure}

By using $n=0.25$, $\Delta_{\rm ex}=0.5$ and $\lambda=0.01$, we can start from a uniform ferromagnetic state ($\bm e_i=\hat{\bm z}$) at $T$=0. Under this condition, we calculate the temperature dependence of the magnetic moment, defined by 
\begin{align}
m_z=-\frac{1}{\pi}{\rm Im}\int_{-\infty}^{\infty}d\epsilon f(\epsilon){\rm Tr}_\sigma \sigma_z \Bar G(\epsilon_+),\tag{13}
\end{align}
together with the thermal average of the exchange field strength normalized by $\Delta_{\rm ex}$ 
\begin{align}
\langle e_z\rangle_{\bm e}=\int w(\bm e)e_z d\bm e.\tag{14}
\end{align}
As shown in Fig. 4, we confirmed that $m_z$ is almost proportional to $\langle e_z\rangle_{\bm e}$, which is a natural feature of the DLM scheme. The $T_{\rm C}$ value was approximately $0.0062W$. This is close to the value expected from the mean-field approximation using $J_0$, which is reasonable because we adopted the single-site approximation.

\begin{figure}
\centering
\includegraphics[width=0.7\textwidth]{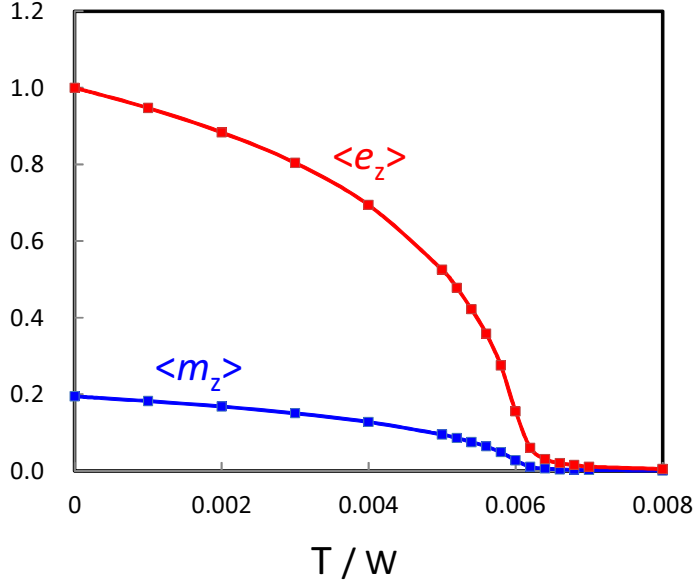}
\caption{The temperature dependence of the magnetic moment $m_z$ and the thermal average of the normalized exchange field strength $\langle e_z\rangle_{\bm e}$ with $\Delta_{\rm ex}=0.5$, $\lambda=0.01$ and $n=0.25$.}
\end{figure}

Figure 5 shows the temperature dependence of $\sigma_{xy}$ for $n=0.25$, $\Delta_{\rm ex}$=0.5, and $\lambda$=0.01. It is worth mentioning here that the Fermi sea term, $\sigma_{xy}^{\amalg}$, is small but finite, whereas in the two-dimensional electron-gas Rashba model, it vanishes at least at $T = 0$, when the Fermi level stays in both branches \cite{Nunner}. This may reflect a characteristic feature of the tight-binding lattice model, where the SOI does not have a simple k-linear dependence. Regarding the Fermi surface term, $\sigma_{xy}^{\rm {I}}$, we must note that both the intrinsic and extrinsic parts increase with increasing temperature. 
Generally, it is expected from the current literature that the intrinsic part $\sigma_{xy}^{\rm {I\,int}}$ exhibits robust or decreasing behaviour against electron scattering, and the extrinsic part $\sigma_{xy}^{\rm {I\,ext}}$ grows divergently with decreasing scattering rate as the longitudinal conductivities do.  
We note here that as far as $\sigma_{xy}^{\rm {I\,int}}$ is concerned, the present behaviour might be 
understood from the $\Delta_{\rm ex}$ dependence of the AHC in a pure system of the Rashba model. 
The AHC in the pure system $(T = 0)$ is described by $\sigma_{xy}=-\frac{e^2}{\hbar}\frac{1}{\Omega}\sum_{\pm,\bm k}f(\epsilon_{\bm k, \pm})B_z^{\pm}(\bm k)$ where $\epsilon_{\bm k, \pm}=-2t(\cos ak_x+\cos ak_y)\pm\{\lambda^2(\sin ^2ak_x+\sin^2ak_y)+\Delta_{\rm ex}^2\}^{1/2}$. 
\begin{figure}
\centering
\includegraphics[width=0.7\textwidth]{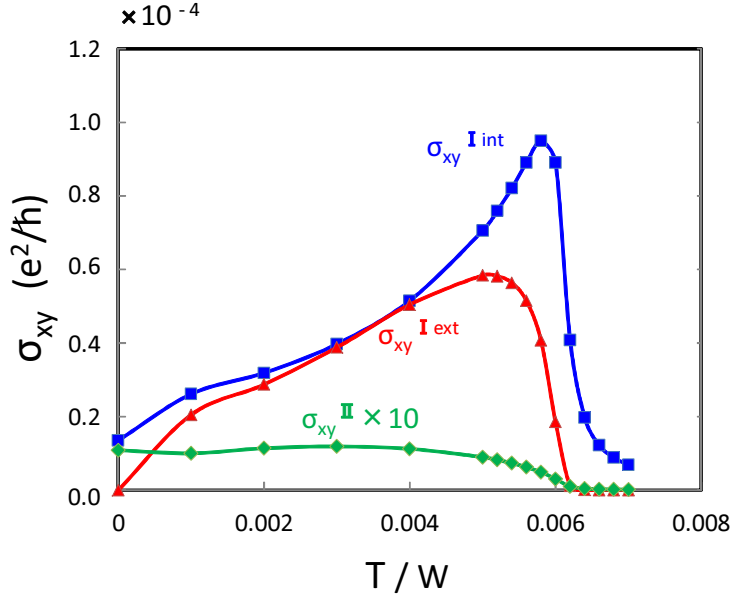}
\caption{The temperature dependences of $\sigma_{xy}^{\rm {I\,int}}$, $\sigma_{xy}^{\rm {I\,ext}}$ and 
$\sigma_{xy}^{\amalg}$ in unit of $e^2/\hbar$, with $\Delta_{\rm ex}=0.5$, $\lambda=0.01$ and $n=0.25$.}
\end{figure}
The Berry curvature $B_z^{\pm}(\bm k)$ is expressed as
\begin{align}
B_z^{\pm}(\bm k)=\pm\frac{1}{2}\frac{\Delta_{\rm ex}\lambda^2a^2\cos ak_x\cos ak_y}{\{\lambda^2(\sin ^2ak_x+\sin^2ak_y)+\Delta_{\rm ex}^2\}^{3/2}}.
\tag{15}\label{eq.15}
\end{align}
Therefore, for $\Delta_{\rm ex}\gg\lambda$, $B_z^{\pm}(\bm k)$ is proportional to $\lambda^2/\Delta_{\rm ex}^2$ and 
$\sigma_{xy}$ approximately leads to $\lambda^2/\Delta_{\rm ex}$ when the two branches (+ and $-$) are partially occupied. At a finite temperature, the coherent potential appearing in eq.(\ref{eq.3a}) is approximately given by $\Sigma(\epsilon)\sim-\Delta_{\rm ex}\langle e_z\rangle_{\bm e}\sigma_z +\Bar{G}(\epsilon)\Delta_{\rm ex}^2(1-\langle e_z\rangle_{\bm e}^2)$. One finds that the real part of $\Sigma(\epsilon)$ is mainly governed by $-\Delta_{\rm ex}\langle e_z\rangle_{\bm e}\sigma_z$, at least when $\langle e_z\rangle_{\bm e}$ is close to unity (low $T$). Then the exchange splitting $\Delta_{\rm ex}$ in eq.(\ref{eq.15}) is approximately replaced by $\Delta_{\rm ex}\langle e_z\rangle_{\bm e}$ at $T>0$. Since the effective exchange splitting $\Delta_{\rm ex}\langle e_z\rangle_{\bm e}$ decreases with temperature as shown in Fig. 4, we can deduce that the intrinsic contribution of the AHC increases with increasing temperature. Standing on this viewpoint, $\sigma_{xy}^{\rm {I\,int}}$ is expected to exhibit a maximum value when $\Delta_{\rm ex}\langle e_z\rangle_{\bm e}\sim\lambda$ is satisfied. This leads to $\langle e_z\rangle_{\bm e}\sim0.02$ in our case, and then the peak position of $\sigma_{xy}^{\rm {I\,int}}(T)$ is expected to be $T/W\sim 0.006$ from Fig. 4. One can see that this is consistent with the behaviour in Fig. 5. The above feature can be understood more intuitively if one rewrites the Berry curvature (eq.(\ref{eq.15})) in the form 
\begin{align}
B_z^{\pm}(\bm k)={\pm}\frac{1}{2S_{\bm k}^3}{\bm S_{\bm k}}\cdot(\partial_{k_x}{\bm S_{\bm k}}\times\partial_{k_y}{\bm S_{\bm k}}), \tag{16}\label{eq.16} 
\end{align}
where ${\bm S_{\bm k}}\equiv(-\lambda\sin{ak_y}, \lambda\sin{ak_x}, \Delta_{\rm ex})$.  The vector ${\bm S_{\bm k}}$ can be regarded as a spin texture in the k-space, since the Hamiltonian with EXS in the $z$-direction can be written in the form $H\{\hat{\bm z}\} = -2t\sum_{\bm k,\sigma}(\cos{ak_x}+\cos{ak_y}) n_{\bm k,\sigma} - \sum_{\bm k}\bm S_{\bm k}\cdot\bm {\sigma}_{\bm k}$ where $\bm {\sigma}_{\bm k}=\sum_{\sigma,\sigma'}(\bm {\sigma})_{\sigma,\sigma'} c_{\bm k,\sigma}^{\dagger}c_{\bm k,\sigma'}$. Therefore, one can see that eq.(\ref{eq.16}) corresponds to the spin chirality in the k-space and is strongly dependent on $\Delta_{\rm ex}$ as schematically shown in Fig. 6. This leads us to recognize the physical picture of how $\Delta_{\rm ex}$ affects the AHC through the change of spin chirality in the k-space. In contrast, the intra-atomic SOI in typical 3d systems involves the diagonal component of the spin operator $(l_zs_z)$ in multi-orbital, and this term $l_zs_z$ is effective to $\sigma_{xy}^{\rm {I\,int}}$  even though the $\Delta_{\rm ex}$ is infinite \cite{Kontani}. Therefore, the intrinsic contribution of the AHC in most transition metal systems is not so sensitive to the exchange splitting $\Delta_{\rm ex}$ or shows a decreasing behaviour with decreasing $\Delta_{\rm ex}$. In this sense, the present $\sigma_{xy}^{\rm {I\,int}}$ is considered to reflect a characteristic feature of Rashba-type ferromagnets satisfying  $\Delta_{\rm ex}\gg\lambda$. Actually in $\Delta_{\rm ex}<\lambda$ case, the above situation does not hold and the behaviour is changed. Figure 7 shows $\sigma_{xy}(T)$ for $\Delta_{\rm ex}=0.1$, $\lambda=0.5$ and $n=0.25$ (the case for Fig. 3), for comparison. One finds that $\sigma_{xy}^{\rm {I\,int}}$ is not so sensitive and almost flat against the temperature change. 

\begin{figure}
\centering
\includegraphics[width=0.7\textwidth]{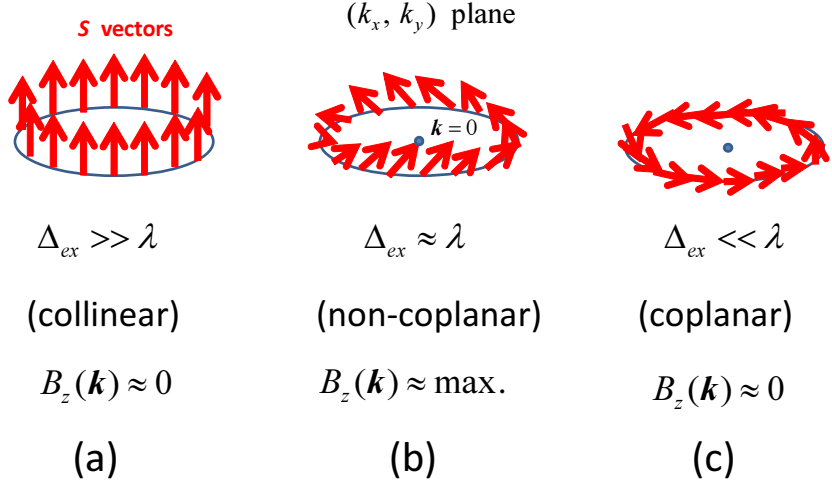}
\caption{Schematic pictures of spin texture ${\bm S_{\bm k}}=(-\lambda\sin{ak_y}, \lambda\sin{ak_x}, \Delta_{\rm ex})$ in the k-space. 
In the case of $\Delta_{\rm ex}\gg\lambda$ (a), collinear structure is realized and then the spin chirality is diminished. For $\Delta_{\rm ex}\sim\lambda$ (b), the spin chirality reaches to maximum, and when $\Delta_{\rm ex}\ll\lambda$ (c), the spin chirality decreases again. 
}
\end{figure}

\begin{figure}
\centering
\includegraphics[width=0.7\textwidth]{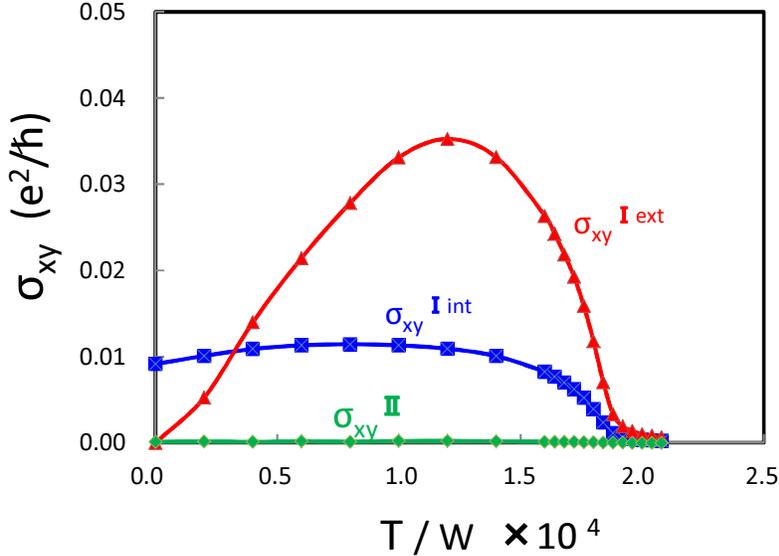}
\caption{The temperature dependences of $\sigma_{xy}^{\rm {I\,int}}$, $\sigma_{xy}^{\rm {I\,ext}}$ and 
$\sigma_{xy}^{\amalg}$ in unit of $e^2/\hbar$, with $\Delta_{\rm ex}=0.1$, $\lambda=0.5$ and $n=0.25$.}
\end{figure}

We should stress here that, as far as the spin Hall conductivity (SHC) is concerned, the intrinsic part is expected to increase with increasing temperature, regardless of the type of SOI.  This is because the decrease of effective $\Delta_{\rm ex}$ always enlarges the spin currents in spin Hall phenomena.  Zhang et al.\cite{Jiang} have recently measured that the SHC of Py(FeNi)/Pt increases with increasing temperature. They attributed this behaviour to the decrement of magnetization (effective exchange splitting) in Pt layer induced by the magnetic proximity effect.  We consider that this experiment indirectly supports the present result and the above explanation.

Additionally, we must refer to the theoretical work by Ye et al.\cite{Ye}, who explained the experimental 
result of the AHC of La$_{0.7}$Sr$_{0.3}$MnO$_3$, a colossal-magnetoresistance material. The authors suggested that the Berry phase due to skyrmions arises with increasing temperature and can induce AHC in the presence of the SOI in double-exchange (half-metallic) ferromagnets. This mechanism also corresponds to an intrinsic origin as well as the present one. However, our treatment is restricted to the single-site approximation and  is therefore not available to identify skyrmions or spin chirality in real space. Instead, the present results reflect the variation of spin chirality (skyrmion density \cite{G. Tatara}) in k-space (eq.(\ref{eq.16})) through the change of effective EXS $\Delta_{\rm ex}\langle e_z\rangle_{\bm e}$. In this sense, the physical feature underlying the mechanism of the enhancement of AHC is common to both cases.

Regarding the extrinsic contribution $\sigma_{xy}^{\rm {I\,ext}}$ in Figs. 5 and 7, it seems that the increasing behaviour with temperature cannot be described in line with previous theories. 
It has been believed that the skew scattering part of $\sigma_{xy}^{\rm {I\,ext}}$ diverges when the impurity scattering rate goes to zero as well as $\sigma_{xx}$, whereas in Figs. 5 and 7, $\sigma_{xy}^{\rm {I\,ext}}$  vanish at $T=0$.  One should note here that the divergent behaviour of the skew scattering part of $\sigma_{xy}^{\rm {I\,ext}}$ in the impurity scattering case is driven by a decrease of impurity concentration ($n_{\rm imp}\rightarrow0$). Thus, in the clean cases, AHC is dominated by the extrinsic part. However, as shown by Onoda et al.\cite{Onoda}, for a smaller impurity potential strength (not $n_{\rm imp}$), the extrinsic part is suppressed and then the AHC is mainly dominated by $\sigma_{xy}^{\rm {I\,int}}$ ($\sigma_{xy}^{\rm {I\,ext}}\sim0$). Based upon this aspect, it is natural for $\sigma_{xy}^{\rm {I\,ext}}$ in the present case to vanish at $T=0$ because of shrinkage of the scattering strength and to increase with increasing temperature owing to growth of the scattering strength. 

Finally, it may be meaningful to refer to the theoretical work of Kondo \cite{Kondo}. He calculated the AH resistivity using the s-d model including the SOI and obtained the result $\rho_{xy}\propto(1-\langle e_z\rangle_{\bm e})^3$. His treatment corresponds to an extrinsic contribution of $\sigma_{xy}$ due to the local spin fluctuation. We suppose that the mechanism suggested by Kondo is essentially the same as the present one for $\sigma_{xy}^{\rm {I\,ext}}$  and can be regarded as a natural feature in typical transition metal 
systems. Note, however, that in actual systems having multi-orbital with intra-atomic SOI, the vertex correction terms ($\sigma_{xy}^{\rm {I\,ext}}$) vanish or have a small contribution to the AHC when the relevant system has an almost 3d character~\cite{Tanaka}, because, in this case, the velocity vertex parts (eq.(\ref{eq.7a})) are composed of odd parity. In contrast, in the Rashba model, eq.(\ref{eq.7a}) includes even parity owing to the Rashba-type SOI. Then, the scattering event appears more effectively in the vertex correction term in the Rashba model, resulting in a relatively large contribution to $\sigma_{xy}^{\rm {I\,ext}}$.

\section{Summary}
We investigated the AHC of Rashba-type ferromagnets using the tight-binding lattice model at finite temperature considering spin fluctuation. 
The most distinctive feature we observed is that the intrinsic AHC increases with increasing temperature. This can be understood from the perspective of Berry curvature at $T=0$, which indicates that the AHC increases with decreasing EXS when the SOI is much smaller than the EXS. Qualitatively, this can be linked to the spin chirality in the k-space which is maximum for $\Delta_{\rm ex}\sim\lambda$ (non-coplanar spin structure) and is diminished both for $\Delta_{\rm ex}\gg\lambda$ (collinear spin structure) and $\Delta_{\rm ex}\ll\lambda$ (coplanar spin structure). The extrinsic part of the Fermi surface term also increases with increasing temperature starting from 0 at $T=0$ and has a large contribution at finite temperatures, comparable to the intrinsic part. This seems contradictory behaviour to the usual skew scattering case in which $\sigma_{xy}^{\rm {I\,ext}}$ grows divergently when the impurity concentration goes to zero.  However, the present result is considered to be natural because the temperature change does not imply concentration change of scattering centres but instead corresponds to variation of the scattering strength.

In principle, the single-site approximation employed here for the spin fluctuations is not appropriate to the two-dimensional system and the results may not reach the quantitative level. However, if a ferromagnetic state is realized in an actual bi-layer system, we believe that the physical pictures found in this work might lurk as an AHE in a system where the Rashba-type SOIs exist.
    
\section{Acknowledgement}
The author wishes to thank to Professor J. I. Inoue for useful discussion. This work was supported 
by JSPS Kakenhi (Grant No. 16K06702) and CSRN in Japan.

\end{document}